\documentclass[twocolumn,preprintnumbers,amsmath,amssymb,floatfix]{revtex4}

\usepackage{graphicx}
\usepackage{dcolumn}
\usepackage{bm}
\usepackage{color}
\def\degree{{\mathrm o}}

\usepackage{hyperref}

\usepackage{latexsym,epsfig,rotate} 

\usepackage{cancel} 

\newcommand{\be}{\begin{equation}}
\newcommand{\ee}{\end{equation}}

\def\bge{\begin{equation}}
\def\ene{\end{equation}}
\def\bgea{\begin{eqnarray}}
\def\enea{\end{eqnarray}}
\def\nn{\nonumber}
\def\imnim{i}

\def\ls{\raise 1.5pt\hbox{$\,<\;$}\kern -10.5pt\lower3.5pt
          \hbox{$\sim$}\kern 1.5pt} 
\def\gs{\raise 1.5pt\hbox{$\,>\,$}\kern -9.5pt\lower3.5pt
          \hbox{$\sim$}\kern 1.5pt} 

\begin{document}
\frenchspacing 
\preprint{IFIC/11-63}
\preprint{FTUV/2011-1110}

\title{Two-photon form factors of the $\pi^0$, $\eta$ and $\eta^\prime$ mesons 
       in the chiral theory with resonances}
\sloppy
\author{Henryk Czy\.z}
\affiliation{%
Institute of Physics, University of Silesia, Katowice PL-40007, Poland
}%
\author{Sergiy Ivashyn, Alexandr Korchin}
\affiliation{%
NSC ``Kharkov Institute of Physics and Technology'', Kharkov UA-61108, Ukraine
}%
\author{Olga Shekhovtsova}
\affiliation{%
IFIC, Universitat de Val\`encia-CSIC, Apt.~Correus 22085, E-46071 Val\`encia, Spain
}%

\date{\today}

\begin{abstract}
 We have developed a phenomenological approach which 
 describes very well the $\pi^0$, $\eta$ and $\eta^\prime$ meson
 production in the two-photon interactions.
 The simultaneous description of the $\pi^0$, $\eta$ and $\eta^\prime$ 
 meson two-photon form factors is consistent with data in the space-like region. 
 The obtained form factors are implemented in the event generator EKHARA
 and the simulated cross sections are presented.
 Uncertainties in the measured form factors coming 
 from the model dependence in Monte Carlo simulations are studied. 
 The model predictions for the form factor slopes at the origin
 are given and the high-$Q^2$ limit is also discussed.
\end{abstract}


\maketitle

\section{Introduction}

The two-photon transition form factors of the pseudoscalar mesons
$\pi^0$, $\eta$ and $\eta^\prime$
have received a great attention lately -- both from the experimental 
and the theoretical side.
The recent BaBar experiments~\cite{Aubert:2009mc,:2011hk} have
provided us with important information in the high-$Q^2$ region
of the photon virtuality and have triggered new insight 
into the structure of mesons~\cite{Dorokhov:2009jd,Kroll:2010bf,Brodsky:2011xx,Brodsky:2011yv,Klopot:2011qq,Balakireva:2011wp,Bakulev:2011rp,Dorokhov:2011dp}.
Hopefully, results from Belle experiment will soon be available,
and will provide a very important cross-check of the BaBar  
data and boost a progress in the form factor phenomenology.
A new experiment KLOE-2 at Frascati~\cite{AmelinoCamelia:2010me,Babusci:2011bg} will soon
be able to provide us with the information on the pion 
two-photon form factor at low $Q^2$ -- in a region where no measurements
were available~\cite{Babusci:2011bg}.
Also the transition form factors of $\mathcal{P} = \pi^0,\eta,\eta^\prime$
(and other) mesons will be measured in BES-III~\cite{Asner:2008nq}
experiment at Beijing with high statistics.
 
The Monte Carlo generators based on reliable models
are needed for data analysis and feasibility studies.
One of the tools in this field is the Monte Carlo generator 
EKHARA~\cite{Czyz:2010sp,Czyz:2006dm},
which is already in use by KLOE-2 Collaboration~\cite{Babusci:2011bg}.
A realiable simulation has to account for
both photon virtualities in the form factor
even for a ``single-tag'' experiment.
Therefore, the formulae for the form factors as functions 
of two photon virtualities are needed.
This criterion considerably reduces the choice
for the form factor,
because the majority of the published formulae 
within different theoretical approaches 
hold only for the case with one photon being real and the other
--- space-like and virtual.

It is worthwhile to stress that the knowledge of the transition
form factors is important in itself, but it is also required
for the calculation of the hadronic light-by-light (hLbyL) scattering
part of the anomalous magnetic moment of 
the muon ($a_\mu^{\mathrm{LbyL}}$), see, 
e.g.,~\cite{Melnikov:2003xd,Prades:2009tw,Nyffeler:2009tw,Jegerlehner:2009ry}.

In order to take full advantage from  the newly planned $g-2$
experiments at Fermilab~\cite{Carey:2009zzb} and JPARC~\cite{JParc:amm:2010}, 
 it is mandatory to improve the accuracy of the hLbyL contribution. 
This subject has been recently discussed in detail during the dedicated
workshop in Seattle (\href{http://www.int.washington.edu/PROGRAMS/11-47w/}{{\tt http://www.int.washington.edu/PROGRAMS/11-47w/}}).
Many important issues related to
the $\gamma^\ast\gamma^\ast\mathcal{P}$ interaction
have recently been discussed in~\cite{Bernstein:2011bx}.

It has not been feasible so far to develop a rigorous 
QED/QCD based theoretical description of the two-photon interaction of mesons,
which would be applicable at an arbitrary energy scale.
Various methods have been used, depending on the aim of a research:
the Brodsky-Lepage (BL) 
high-$Q^2$ limit and interpolation formula~\cite{Lepage:1980fj};
the Operator Product Expansion (OPE) approach 
to Vector-Vector-Pseudoscalar (VVP) and Vector-Vector-Axial (VVA)
three-point functions of QCD~\cite{Knecht:2001xc};
the Vector Meson Dominance (VMD) models~\cite{Silagadze:2006rt,Lichard:2010ap};
the holographic QCD approaches~\cite{Grigoryan:2008up,Stoffers:2011xe,Cappiello:2010uy,Brodsky:2011xx};
the QCD sum rules~\cite{Khodjamirian:1997tk,Agaev:2010aq,Klopot:2011qq,Balakireva:2011wp,Bakulev:2011rp};
the modified perturbative approach~\cite{Kroll:2010bf};
the Regge models~\cite{Arriola:2010aq};
the Dyson-Schwinger equation~\cite{Roberts:2010rn};
the Nambu-Jona-Lasinio model~\cite{Bartos:2001pg,Noguera:2011fv},
the constituent quark models~\cite{Dorokhov:2009jd,Dorokhov:2011zf};
the Resonance Chiral Theory approach~\cite{Mateu:2007tr,Kampf:2011ty};
and others.
The research in this field is mainly dedicated 
to the high-$Q^2$ region of the form factor with one real
and one virtual photon.
When one needs to cover a wide range of the photon virtuality,
both high-energy and low-energy methods have to be merged in 
some appropriate way.

The purpose of this paper is twofold:
to develop a reliable model able to 
describe the two-photon form factors 
of $\pi^0$, $\eta$ and $\eta^\prime$ mesons 
with a very small number of parameters
and, then, to implement these form factors 
in the generator EKHARA.

Our approach is described in Section~\ref{sec:ffs}.
We start from the formalism of chiral effective theory 
with resonances~\cite{Ecker:1989yg,Ecker:1988te,Prades:1993ys} 
as a phenomenological model.
The masses of the particles are taken from PDG~\cite{Nakamura:2010zzi}, 
and the $\eta$--$\eta^\prime$ mixing is accounted for
according to Ref.~\cite{Feldmann:1998vh,Feldmann:1999uf}.
We require that the form factors vanish at high $|t|$.
  The formulae for the form factors are given in Section~\ref{sec:results}.
  In Sections~\ref{sec:1octets} and~\ref{sec:2octets:fit}
  we compare the calculated form factors
  with the experimental results of 
  CELLO~\cite{Behrend:1990sr}, CLEO~\cite{Gronberg:1997fj} 
  and BaBar~\cite{Aubert:2009mc,:2011hk} experiments.
  Furthermore, in Section~\ref{sec:mc},
  we discuss the implementation of the calculated form factors in 
  the Monte Carlo generator EKHARA~\cite{Czyz:2010sp,Czyz:2006dm}
  and simulate the single-tag visible 
  cross section, under conditions similar to those of 
  the CLEO and BaBar experiments.
  The derived formulae allow us to study 
  the high-$Q^2$ behavior of the form factor
  (Section~\ref{sec:limit:high})
  and the slope of the form factor 
  at the origin (Section~\ref{sec:slope}).
  Finally, in Section~\ref{sec:summary} the main conclusions of the paper
  are drawn. 

\section{Our approach and the results}
\label{sec:ffs}
\subsection{Formulae for $F_{\gamma^\ast\gamma^\ast\mathcal{P}}$}
\label{sec:results}
The two-photon form factor $F_{\gamma^\ast\gamma^\ast \mathcal{P}}(t_1, t_2)$
for the meson of type $\mathcal{P} = \pi^0,\eta,\eta^\prime$,
encodes the dependence of the amplitude 
$\mathcal{M}\left( \gamma^\ast\gamma^\ast \to \mathcal{P} \right)$
on the virtuality of the photons
($q_1^2 = t_1$, $q_2^2 = t_2$):
\bgea
&&\mathcal{M}[\gamma^\ast(q_1, \nu)\, \gamma^\ast(q_2, \beta) \to \mathcal{P}]
 \\ \nn
&& \quad \quad = 
e^2 \epsilon_{\mu\nu\alpha\beta} q_{1}^{\mu} q_{2}^{\alpha} 
F_{\gamma^\ast\gamma^\ast \mathcal{P}}(t_1, t_2)
,
\enea
where $\epsilon_{\mu \nu \alpha \beta}$ is the totally antisymmetric
Levi-Civita tensor.
Note that $F_{\gamma^\ast\gamma^\ast \mathcal{P}}(t_1, t_2) = F_{\gamma^\ast\gamma^\ast \mathcal{P}}(t_2, t_1)$ due to Bose symmetry of the photons.
We obtain the formulae for the form factors 
$F_{\gamma^\ast\gamma^\ast\mathcal{P}}(t_1,t_2)$ 
on the basis of the effective chiral 
Lagrangian~\cite{Ecker:1989yg,Ecker:1988te,Prades:1993ys}
extended to multi-octet resonance contributions,
with the $\eta-\eta'$ mixing accounted for
as in~\cite{Feldmann:1998vh,Feldmann:1999uf}.
A brief summary of the model is given in Appendix~\ref{sec:formalism}.
 We would like to remark that
 a similar approach was applied in the context of other
 processes in~\cite{Ivashyn:2007yy,Ivashyn:2009te,Eidelman:2010ta}.  

\begin{figure}
\begin{center}
\resizebox{0.5\textwidth}{!}{%
     \includegraphics{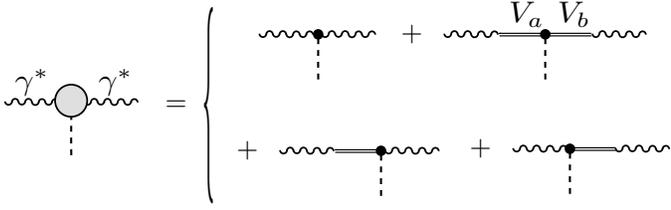}
     }
\end{center}
\caption{Diagrams for the $\gamma^*\gamma^*\mathcal{P}$ transition.
Dashed lines correspond to the pseudoscalar meson $\mathcal{P}$,
solid lines --- to the vector mesons 
and wavy lines to (virtual) photons.
$V_a \ne V_b$ for $\pi^0$
and $V_a = V_b$ for $\eta$ and $\eta^\prime$
form factors.}
\label{fig:gamgampi}
\end{figure}

For simplicity we neglect the mixing between the
octets, which can be added if required by the data.
The diagrams describing $\gamma^\ast\gamma^\ast P$ transition 
are presented in Fig~\ref{fig:gamgampi}. 
The form factors read
\begin{widetext}
\bgea\label{eq:ffpi_3octet}
&& F_{\gamma^\ast\gamma^\ast\pi^0}(t_1,t_2)=-\frac{N_c}{12\pi^2 f_\pi}
+\sum_{i=1}^{n}\frac{4\sqrt{2}h_{V_i} f_{V_i}}{3f_\pi} {t_1} \biggl({D_{\rho_i} (t_1)}
+{D_{\omega_i} (t_1) }\biggr) \\
&&+\sum_{i=1}^{n}\frac{4\sqrt{2}h_{V_i} f_{V_i}}{3f_\pi} {t_2} \biggl({D_{\rho_i} (t_2) }
+{D_{\omega_i} (t_2)}\biggr)-\sum_{i=1}^{n}\frac{4\sigma_{V_i} f_{V_i}^2}{3f_\pi} {t_2}  {t_1}
\biggl({D_{\rho_i} (t_2) }{D_{\omega_i} (t_1) } +{D_{\rho_i} (t_1) } {D_{\omega_i} (t_2)
} \biggr) ,\nonumber
\enea
\bgea\label{eq:ffeta_3octet}
&& F_{\gamma^\ast\gamma^\ast\eta}(t_1,t_2)= 
-\frac{N_c}{12\pi^2f_\pi}
\left( \frac{5}{3} C_q  -  \frac{\sqrt{2}}{3} C_s\right)
\\
&&+\sum_{i=1}^{n}\frac{4\sqrt{2}h_{V_i} f_{V_i}}{3f_\pi} t_1
 \biggl(3 C_q{D_{\rho_i} (t_1)}
+\frac{1}{3}C_q{D_{\omega_i} (t_1)}
-\frac{2\sqrt{2}}{3}C_s{D_{\phi_i} (t_1)}\biggl) 
\nonumber
\\
&&+\sum_{i=1}^{n}\frac{4\sqrt{2}h_{V_i} f_{V_i}}{3f_\pi} t_2 \biggl(3 C_q{D_{\rho_i}
(t_2)}
+\frac{1}{3}C_q {D_{\omega_i}(t_2)} 
- \frac{2\sqrt{2}}{3} C_s {D_{\phi_i} (t_2)}
\biggl)
 \nonumber \\
&& 
-\sum_{i=1}^{n}\frac{8\sigma_{V_i} f_{V_i}^2}{f_\pi} t_2 t_1 \biggl(
\frac{1}{2} C_q{D_{\rho_i} (t_2)} {D_{\rho_i}(t_1)}
 + \frac{1}{18} C_q {D_{\omega_i}(t_2)} {D_{\omega_i} (t_1)} 
 - \frac{\sqrt{2}}{9} C_s {D_{\phi_i} (t_2)} {D_{\phi_i} (t_1)}
\biggl)  \, .
 \nonumber \\
\nonumber \\
&&F_{\gamma^\ast\gamma^\ast\eta^\prime}(t_1,t_2) = 
F_{\gamma^\ast\gamma^\ast\eta}(t_1,t_2)
{}_{\left|
\begin{array}{cr}
 C_q \to &C_q^\prime\\
 C_s \to &-C_s^\prime\\
\end{array}
\right.} , 
\label{eq:ffetapr_3octet}
\enea
\end{widetext}
where $n$ is a number of the vector meson resonance octets.
The definitions of all couplings can be found in the Appendix~\ref{sec:formalism}.
The vector meson propagators $D_V$ are
\bgea
\label{vector-propagator-simple} D_V(Q^2) &= &[Q^2 - M_V^2 + \imnim
\sqrt{Q^2} \Gamma_{tot, V} (Q^2)]^{-1} .
\enea
In this paper we consider only the data in the space-like 
region of photon virtuality, thus the modeling of 
the vector resonance energy dependent widths
$\Gamma_{tot, V} (Q^2)$ is not relevant as the widths are equal to zero. 
We take the values of the masses of all particles
according to PDG~\cite{Nakamura:2010zzi}.

We require that the form factors $F_{\gamma^\ast\gamma^\ast\mathcal{P}}(t_1,t_2)$ 
given in~(\ref{eq:ffpi_3octet}), (\ref{eq:ffeta_3octet}) 
and (\ref{eq:ffetapr_3octet}) 
vanish when the photon virtuality $t_1$ goes to infinity for 
any value of $t_2$:
\bgea
\Bigl.
\lim_{t_1\rightarrow -\infty}
F_{\gamma^\ast\gamma^\ast\mathcal{P}}(t_1,t_2) \Bigr|_{t_2 = const} &=& 0 .
\label{eq:ff_at_inf:1}
\enea
Notice, that in this case the conditions
\bgea
\lim_{t\rightarrow -\infty}
F_{\gamma^\ast\gamma^\ast\mathcal{P}}(t,t)&=& 0 ,
\label{eq:ff_at_inf:1:pure} \\
\lim_{t\rightarrow -\infty}
    F_{\gamma^\ast\gamma^\ast\mathcal{P}}(t,0)&=& 0 
\label{eq:ff_at_inf:2:pure} 
\enea
are automatically satisfied,
which is considered as a correct short-distance behavior of 
the form factors (see, for example, discussion in~\cite{Knecht:2001xc}). 
The constraint~(\ref{eq:ff_at_inf:1}) 
leads to the following relations for the couplings:
\bgea
&& 
\sqrt{2} h_{V_i} f_{V_i} -  \sigma_{V_i}
f_{V_i}^2 = 0 ,\quad i=1,\ldots,n \, ,
\label{eq:qcdsum:1b} \\
&& -\frac{N_c}{ 4 \pi^2} + 8\sqrt{2} \sum_{i=1}^{n} h_{V_i} f_{V_i} = 0
\label{eq:qcdsum:2b} \, .
\enea
Therefore, 
for an ansatz with $n$ vector resonance octets
the two-photon form factors $F_{\gamma^\ast\gamma^\ast \mathcal{P}}(t_1, t_2)$ 
are determined by $2n$ parameters (i.e., the products of the couplings:
$f_{V_i}h_{V_i}$ and $\sigma_{V_i}f_{V_i}^2$, $i=1,\ldots,n$),
from which $n-1$ are to be determined by experiment 
and the rest $n+1$ are fixed by~(\ref{eq:qcdsum:1b}) and~(\ref{eq:qcdsum:2b}).
For the one octet ansatz there are no free parameters
and in case of the two octets ansatz there is one free parameter.

One of the main objectives of this paper was to develop a reliable model for
 the $\gamma^*\gamma^*\mathcal{P}$ ($\mathcal{P}=\pi^0,\eta,\eta'$) transition form
 factors in the space-like region reflecting the experimental data and
 theoretical constrains and
 in the same time being as simple as possible.
Even if we know that the $SU(3)$ flavor symmetry is broken we start our 
 investigations using an $SU(3)$-symmetric model (apart from the
 masses of the mesons, which are fixed at their
 PDG~\cite{Nakamura:2010zzi} values) and try to see how
 many resonance octets we have to include in order to describe the data well.
The existing data for the transition form factors in space-like 
region~\cite{Behrend:1990sr,Gronberg:1997fj,Aubert:2009mc,:2011hk}
 come from single-tag experiments, where one of the invariants 
 is very close to zero (the one associated with the ``untagged'' lepton), 
 thus we have information only about
 $F_{\gamma^\ast\gamma^\ast\mathcal{P}}(t,0)$.
 It is common to define 
 the $\gamma^*\gamma\mathcal{P}$ form factor 
 $F_{\mathcal{P}}(Q^2,0)\equiv F_{\gamma^\ast\gamma^\ast\mathcal{P}}(t,0)$
 with $Q^2\equiv -t$ (associated with the ``tagged'' lepton). 
From Eqs.~(\ref{eq:ffpi_3octet}),~(\ref{eq:ffeta_3octet}) and~(\ref{eq:ffetapr_3octet})
  we see that $F_{\mathcal{P}}(Q^2,0)$ is driven by 
  $n$ parameters (i.e., the products of the couplings:
  $f_{V_i}h_{V_i}$, $i=1,\ldots,n$) 
and there is always only one constraint~(\ref{eq:qcdsum:2b}) for any $n$.
Therefore, the number of parameters in $F_{\mathcal{P}}(Q^2,0)$
to be determined by experiment (``free parameters'') equals to $n-1$ 
(similarly to the case of the $F_{\gamma^\ast\gamma^\ast\mathcal{P}}(t_1,t_2)$).
In case of the one octet ansatz there are no free parameters
and in the two octets case there is one free parameter.

\subsection{The one octet ansatz for the form factors}
\label{sec:1octets}

\begin{figure}
\begin{center}
\resizebox{0.45\textwidth}{!}{%
     \includegraphics{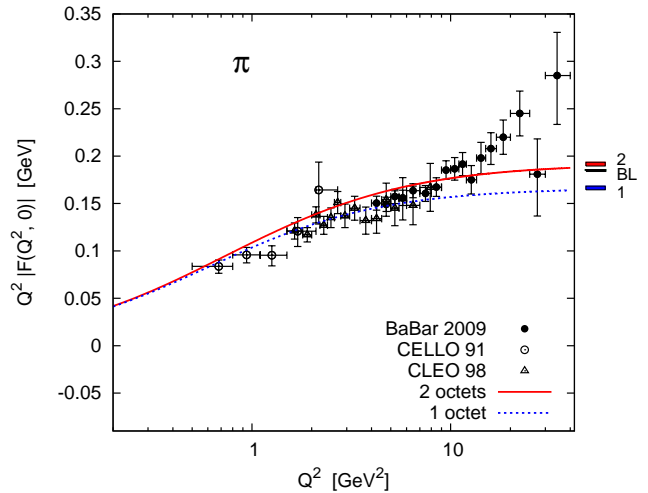}
     }
\end{center}
\caption{Transition form factor $\gamma^\ast \gamma\pi^0$
 compared to the data. 
 The Brodsky-Lepage~\cite{Lepage:1980fj} high-$Q^2$ limit (BL) 
 is shown as a bold solid straight line at $2\times f_\pi=2\times0.0924$~GeV.
 The high-$Q^2$ limit in our 1 octet ansatz and and 2 octets ansatz
 are marked as (1) and (2), correspondingly.
 }
 \label{fig:ff-pi0}
\end{figure}

\begin{figure}
\begin{center}
\resizebox{0.45\textwidth}{!}{%
     \includegraphics{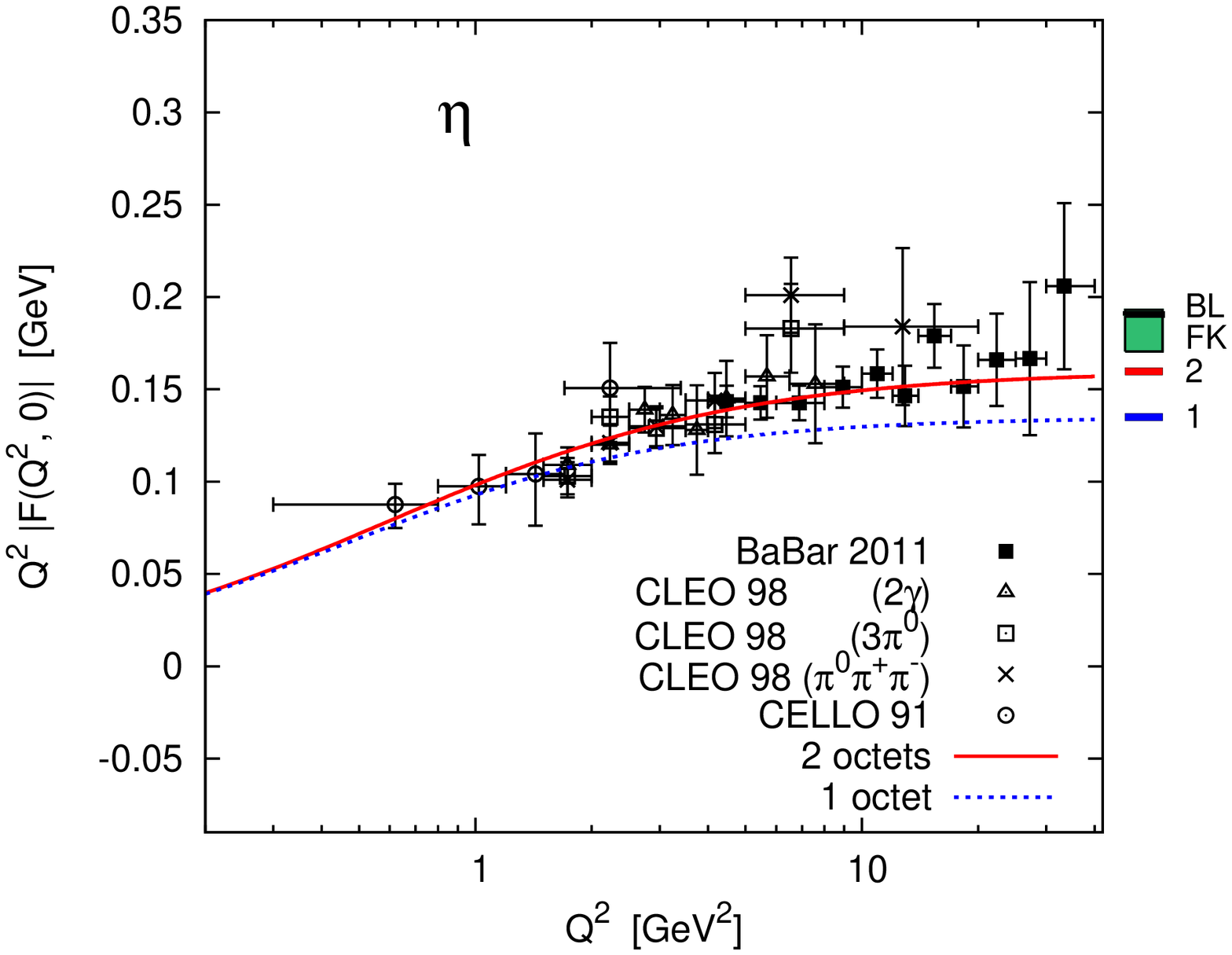}
     }
\end{center}
\caption{Transition form factor $\gamma^\ast \gamma\eta$
 compared to the data.  
 The high-$Q^2$ limit is shown as a bold solid straight 
 line at $2\times f_\eta=2\times0.0975$~GeV,
 according to~\cite{Gronberg:1997fj} and~\cite{Dorokhov:2009jd} (BL).
 The limit according to the two-angle $\eta-\eta^\prime$ 
 mixing scheme~\cite{Feldmann:1998vh,Feldmann:1999uf,Kroll:2010bf} (FK)
 is shown as a shaded box (green online) at $0.1705\ldots0.1931$~GeV,
 accounting for the parameter ambiguities~(\ref{eq:mix_angles}).
 The high-$Q^2$ limit in our 1 octet ansatz and and 2 octets ansatz
 are marked as (1) and (2), correspondingly.
 }
 \label{fig:ff-eta}
\end{figure}

\begin{figure}
\begin{center}
\resizebox{0.45\textwidth}{!}{%
     \includegraphics{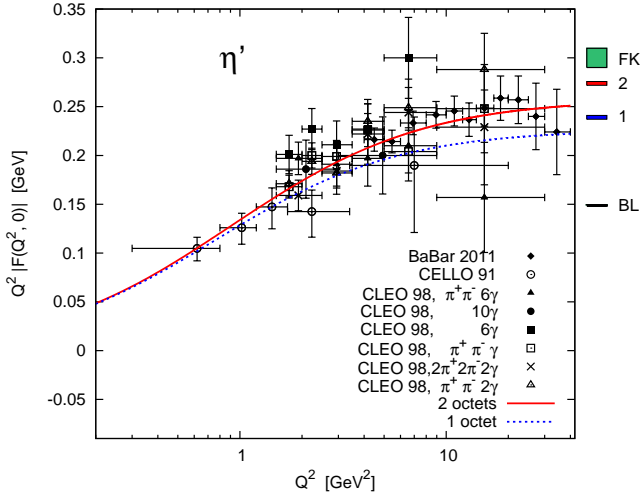}
     }
\end{center}
\caption{Transition form factor $\gamma^\ast \gamma\eta^\prime$
 compared to the data.  
 The high-$Q^2$ limit is shown as a bold solid straight 
 line at $2\times f_{\eta^\prime}=2\times0.0744$~GeV,
 according to~\cite{Gronberg:1997fj} and~\cite{Dorokhov:2009jd} (BL).
 The limit according to the two-angle $\eta-\eta^\prime$ 
 mixing scheme~\cite{Feldmann:1998vh,Feldmann:1999uf,Kroll:2010bf} (FK)
 is shown as a shaded box (green online) at $0.29\ldots0.31$~GeV,
 accounting for the parameter ambiguities~(\ref{eq:mix_angles}).
 The high-$Q^2$ limit in our 1 octet ansatz and and 2 octets ansatz
 are marked as (1) and (2), correspondingly.
 }
 \label{fig:ff-etaP}
\end{figure}

Let us consider first the one octet ansatz. 
In this case 
\bgea
f_{V_1} h_{V_1}= \frac{3}{32\pi^2 \sqrt{2}} ,
\enea
and the model gives a prediction for the form factors 
$F_{\mathcal{P}}(Q^2,0)$
without any possibility for adjustment.
 The predictions of this model are compared with experimental
 data 
 in Figs.~\ref{fig:ff-pi0}-\ref{fig:ff-etaP} (dotted line). 
 To quantify the quality of the agreement of the model predictions 
 we have calculated the $\chi^2$ values for each data set.
  For the pion transition form factor the model agrees with
  CELLO~\cite{Behrend:1990sr} and CLEO~\cite{Gronberg:1997fj} 
  and disagrees with the BaBar data~\cite{Aubert:2009mc},
  as can be seen from Table~\ref{table:chi2}, which shows 
  the $\chi^2$ values per experiment.
  For the $\eta$ and $\eta^\prime$ transition form factor the model 
  is in a perfect agreement with CELLO,
  however for CLEO and BaBar the $\chi^2$ is not good.
  In total, for the one octet ansatz
  we obtain $\chi^2 \approx 358$ for 116 experimental points.

  Even though the overall agreement of this simple model with the data is not
  bad, there is a way to improve it, as will be discussed below.
%

 
\subsection{The two octet ansatz for the form factors}
\label{sec:2octets:fit}

  In order to make the model more flexible,
  we include the second vector meson multiplet contributions.
  We would like to note that there are many known cases
  when in order to improve the model predictions 
  one needs to account for the excited vector resonances,
  the charged form factor of the pion is among the most famous examples.
  
  In the two octet ansatz we chose $h_{V_1}$ as a free parameter
  and determine the value of $f_{V_1}=0.20173(86)$ using the PDG~\cite{Nakamura:2010zzi} 
  value for the width
\bge
\label{eq:fv:pdg}
\Gamma(\rho\to ee) = \frac{e^4 M_\rho f_{V_1}^2}{12\pi}\, .
\ene
  
  The fit to the data gives $\chi^2 \approx 140$ for 116 experimental points.
  The obtained value of $h_{V_1}$ is
  \bgea 
  \label{eq:hv:fit}
  h_{V_1} = 0.03121(14) \, ,
  \enea
  where the error is the parabolic error given by {\tt MINOS} package from
  {\tt MINUIT CERNLIB} program. 
  The remaining coupling is given by
  \bgea
  h_{V_2}f_{V_2}=  \frac{3}{32\pi^2\sqrt{2}}-h_{V_1}f_{V_1}
  = 0.42(5) \times10^{-3} \, .
  \enea
   The comparison of the two octet ansatz with the data is also shown 
   in Figs.~\ref{fig:ff-pi0}-\ref{fig:ff-etaP} (solid line)
   and the $\chi^2$ values per experiment are given in Table~\ref{table:chi2}.
 The only data sample which is not in consistency with the model
 is the BaBar data for $\pi^0$~\cite{Aubert:2009mc}
 (however, for the $\eta$ and $\eta^\prime$ transition form factors
 there is a perfect agreement with BaBar data~\cite{:2011hk}).
 From the plots in Figs.~\ref{fig:ff-pi0}-\ref{fig:ff-etaP}
 and given the numbers in Table~\ref{table:chi2} we conclude
 that the two octet calculation is consistent with the bulk 
 of available data.
%

\begin{table}
\caption{The $\chi^2$ per experiment and the total $\chi^2$.
Number of data points (n.d.p.) is also given for each experiment.
In all given experiments the pseudoscalar meson is produced
in a two-photon process $e^+e^- \to e^+e^-\mathcal{P}$, but 
the decay channels for $\mathcal{P}$ identification vary.
The ``2 octets'' column is calculated with 
the parameter values given by the global fit. }
\label{table:chi2}
\begin{ruledtabular}
\begin{tabular}{ldd}
experiment & \multicolumn{1}{c}{1 octet} & \multicolumn{1}{c}{2 octets} \\
           & \multicolumn{1}{c}{$\chi^2/n.d.p.$}  & \multicolumn{1}{c}{$\chi^2/n.d.p.$} \\
\hline\noalign{\smallskip}
CELLO ($\pi^0\to\gamma\gamma$) 
& 0.29/5 & 0.47/5\\
CLEO  ($\pi^0\to\gamma\gamma$) 
& 6.27/15 & 20.96/15\\
BaBar  ($\pi^0\to\gamma\gamma$) 
& 124.83/17 & 55.85/17\\
\noalign{\smallskip}\hline
CELLO ($\eta\to\gamma\gamma$) 
& 0.24/4 & 0.13/4\\
CLEO  ($\eta\to\pi^+\pi^-\pi^0$) 
& 19.28/6 & 11.13/6\\
CLEO ($\eta\to\gamma\gamma$) 
& 8.55/8 & 2.10/8\\
CLEO  ($\eta\to\pi^0\pi^0\pi^0$) 
& 10.91/5 & 5.63/5\\
BaBar  ($\eta\to\gamma\gamma$) 
& 89.02/11 & 9.34/11\\
\noalign{\smallskip}\hline
CELLO ($\eta^\prime\to\gamma\gamma$) 
& 0.11/5 & 0.29/5\\
CLEO  ($\eta^\prime\to\gamma\gamma\pi^+\pi^-$) 
& 19.90/6 & 7.48/6\\
CLEO  ($\eta^\prime\to\gamma\gamma\pi^+\pi^-\pi^+\pi^-$) 
& 2.61/5 & 1.44/5\\
CLEO  ($\eta^\prime\to\gamma\pi^+\pi^-$) 
& 14.01/6 & 4.64/6\\
CLEO  ($\eta^\prime\to6\gamma$) 
& 21.54/5 & 12.62/5\\
CLEO  ($\eta^\prime\to10\gamma$) 
& 0.49/2 & 0.23/2\\
CLEO  ($\eta^\prime\to\pi^+\pi^-6\gamma$) 
& 5.93/5 & 4.80/5\\
BaBar  ($\eta^\prime\to\gamma\gamma$) 
& 33.87/11 & 3.10/11\\
\noalign{\smallskip}\hline
total & 357.87/116 & 140.22/116\\
\end{tabular}
\end{ruledtabular}
\end{table}

In principle the parameter $h_{V_1}$ can be estimated 
  by experiment via the value of the width
\bge
\label{eq:hv:pdg}
\Gamma(\rho^0\to \pi^0\gamma) = 
\frac{4\alpha M_\rho^3 h_{V_1}^2}{27 f_\pi^2}\left(1-\frac{m_\pi^2}{M_\rho^2}\right)^3 .
\ene
Using the PDG~\cite{Nakamura:2010zzi} values for the width~(\ref{eq:hv:pdg}),
one obtains 
\bge
\label{eq:hv:pdg:value}
h_{V_1} = 0.041(3) \ ,
\ene
which is in tension with the value given by fit~(\ref{eq:hv:fit}).
This might be a result of neglecting the higher octets 
or omission of the $SU(3)$ flavor breaking effects.
In order to check this we have added the third octet to 
the model and fitted two free parameters: 
$h_{V_1}$ and $h_{V_2}f_{V_2}$. 
The fit to the experimental data gives 
$\chi^2 \approx 136$ for~116 experimental points,
so there is no essential improvement 
in the description of data by the model.
   The fit gives $h_{V_1} = 0.03279(75)$ and $h_{V_2}f_{V_2} = -0.73(54)\times10^{-3}$.
   Finally, for the couplings of the third octet one gets 
   $h_{V_3}f_{V_3} = \frac{3}{32\pi^2\sqrt{2}}-h_{V_1}f_{V_1}-h_{V_2}f_{V_2}
    \approx 7.45 \times 10^{-3}$.
   In this fit we observe a very high correlation
   between the parameters $h_{V_1}$ and $h_{V_2}f_{V_2}$ 
   with the off-diagonal correlation coefficient equal to $-0.99$.
   Notice that $h_{V_1}$ is almost unchanged as compared to~(\ref{eq:hv:fit})
   and we conclude that in order to accommodate the
   value~(\ref{eq:hv:pdg:value}) in this model we would need to allow for couplings
   which break the $SU(3)$ flavor symmetry. 
   This is however beyond the scope of the present paper. 
   We leave the possible refinements of the model for further investigations.

In context of the discrepancy between~(\ref{eq:hv:fit}) 
and~(\ref{eq:hv:pdg:value}) we would like to illustrate
the actual experimental uncertainty in the $\rho$ meson decay width.
In Table~\ref{tab:rho_pi_gamma} we show the values of $h_{V_1}$
which are deduced from different experimental values of the width:
the PDG constrained fit~\cite{Nakamura:2010zzi} 
for~$\Gamma(\rho^0\to \pi^0\gamma)$;
the SND measurement~\cite{Achasov:2003ed} 
for~$\Gamma(\rho^0\to \pi^0\gamma)$;
the PDG constrained fit and average~\cite{Nakamura:2010zzi} 
for~$\Gamma(\rho^+\to \pi^-\gamma)$.

\begin{table}
\caption{ The $\rho\to \pi\gamma$ decay width uncertainty
          and the corresponding values of $h_{V_1}$.
        }
\label{tab:rho_pi_gamma}
\begin{ruledtabular}
\begin{tabular}{llll}
\multicolumn{1}{c}{decay} & \multicolumn{1}{c}{width} &
\multicolumn{1}{c}{reference} & \multicolumn{1}{c}{$h_{V_1}$} 
\\
\hline\noalign{\smallskip}
$\rho^0\to \pi^0\gamma$ & $89(12)$~keV & PDG~\cite{Nakamura:2010zzi} &  $0.041(3)$ \\
$\rho^0\to \pi^0\gamma$ & $77(20)$~keV & SND~\cite{Achasov:2003ed} &  $0.038(5)$ \\
$\rho^+\to \pi^+\gamma$ & $68(7)$~keV  & PDG~\cite{Nakamura:2010zzi} &  $0.036(2)$ \\
\end{tabular}
\end{ruledtabular}
\end{table}

\subsection{The Monte Carlo simulation}
\label{sec:mc}

We have implemented  the transition form factors obtained within
the two octet model described above 
into the Monte Carlo generator EKHARA
(\href{http://prac.us.edu.pl/\%7Eekhara}{{\tt http://prac.us.edu.pl/\~\;\!ekhara}}).
From the technical point of view of the event generation, 
it is a straightforward generalization because 
the mappings used in~\cite{Czyz:2010sp} for $\pi^0$
work similarly well also for $\eta$ and $\eta^\prime$.

\begin{figure*}
\begin{center}
\resizebox{0.48\textwidth}{!}{%
     \includegraphics{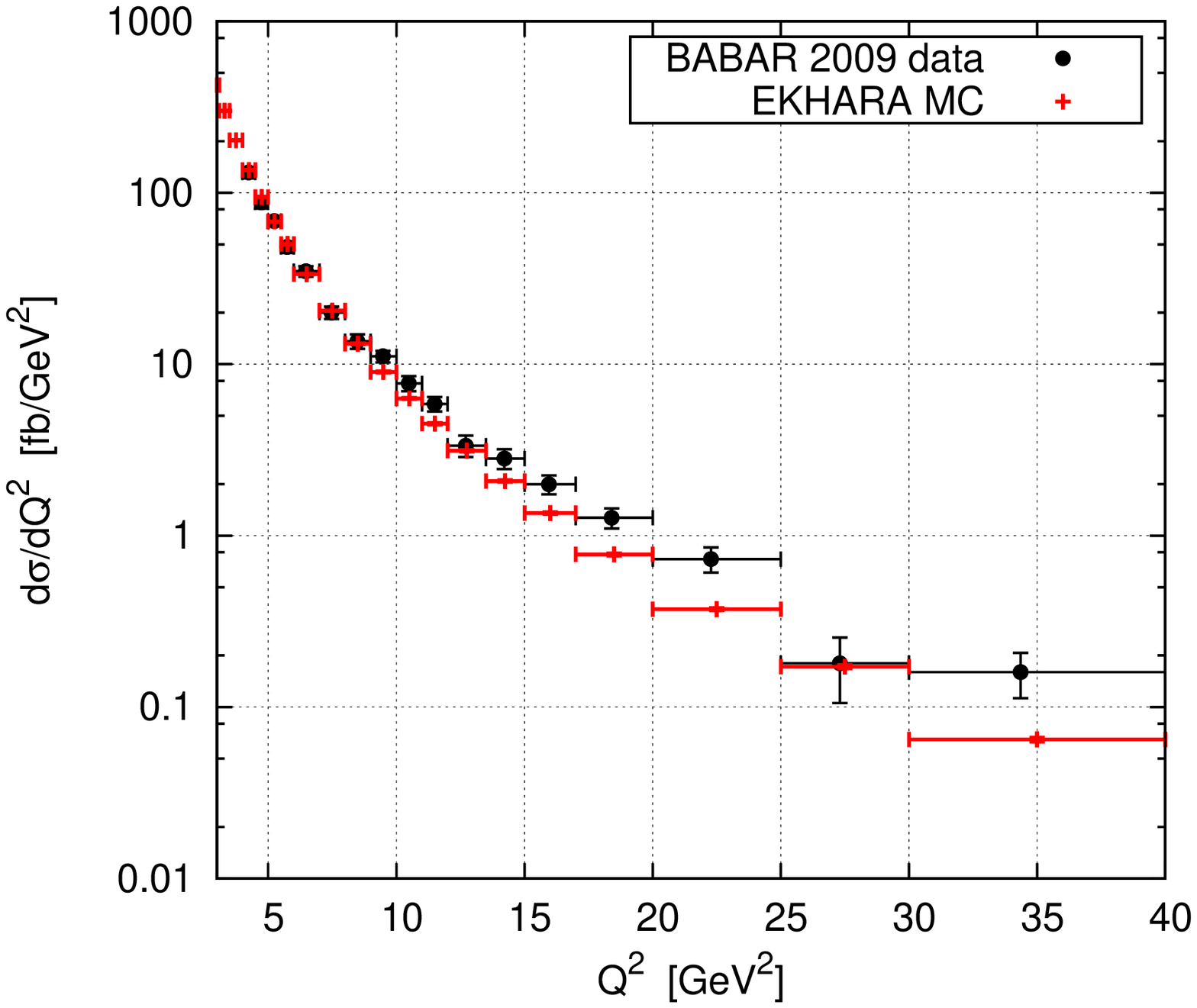}
     }
\resizebox{0.48\textwidth}{!}{%
     \includegraphics{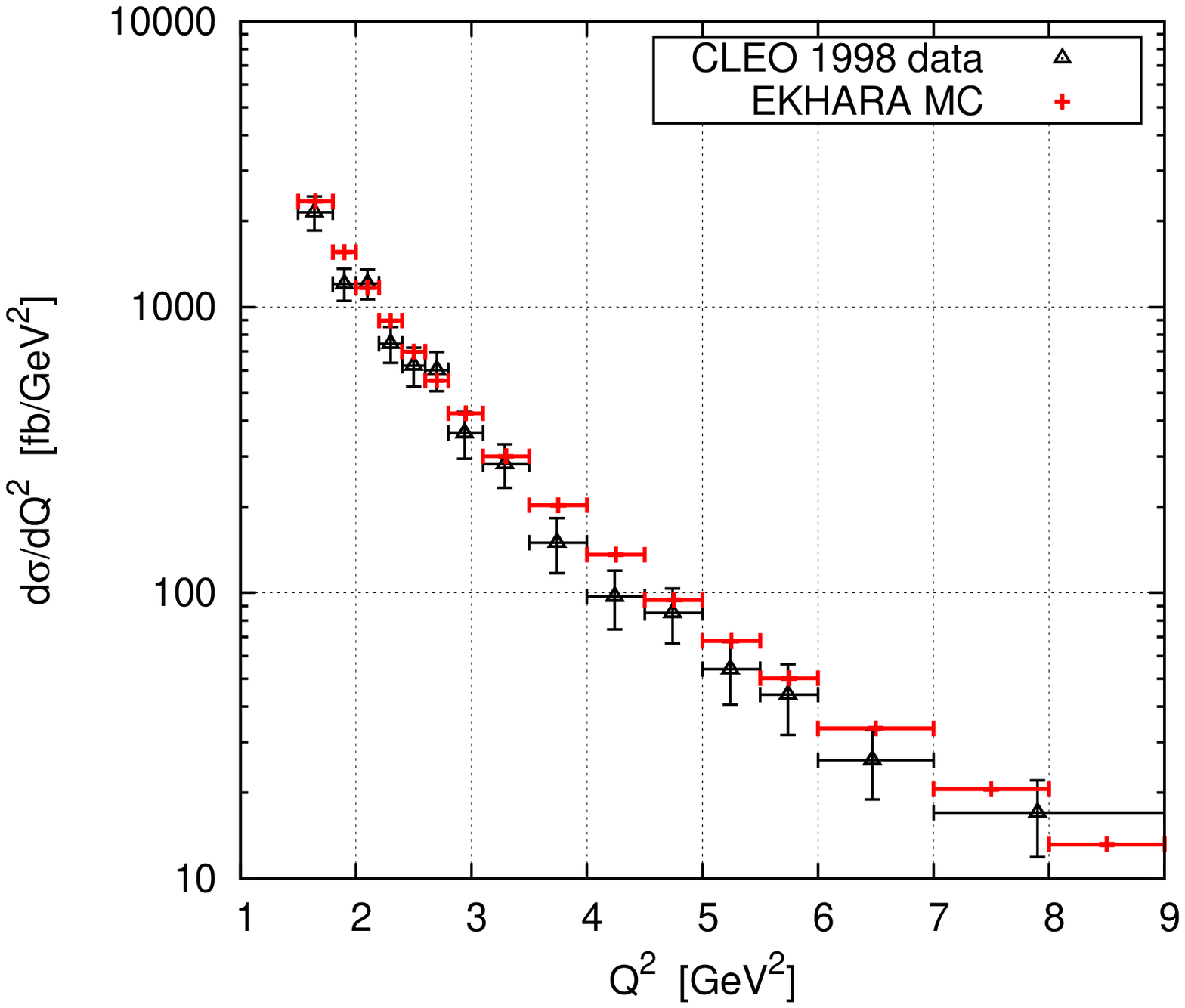}
     }
\end{center}
\caption{The cross section ${d\sigma}/{d Q^2}$
for the process $e^+e^- \to e^+e^-\pi^0$
 compared to BaBar~\cite{Aubert:2009mc} (left)
 and
 CLEO~\cite{Gronberg:1997fj} (right). 
 }
 \label{fig:cs-pi0}
\end{figure*}

\begin{figure*}
\begin{center}
\resizebox{0.48\textwidth}{!}{%
     \includegraphics{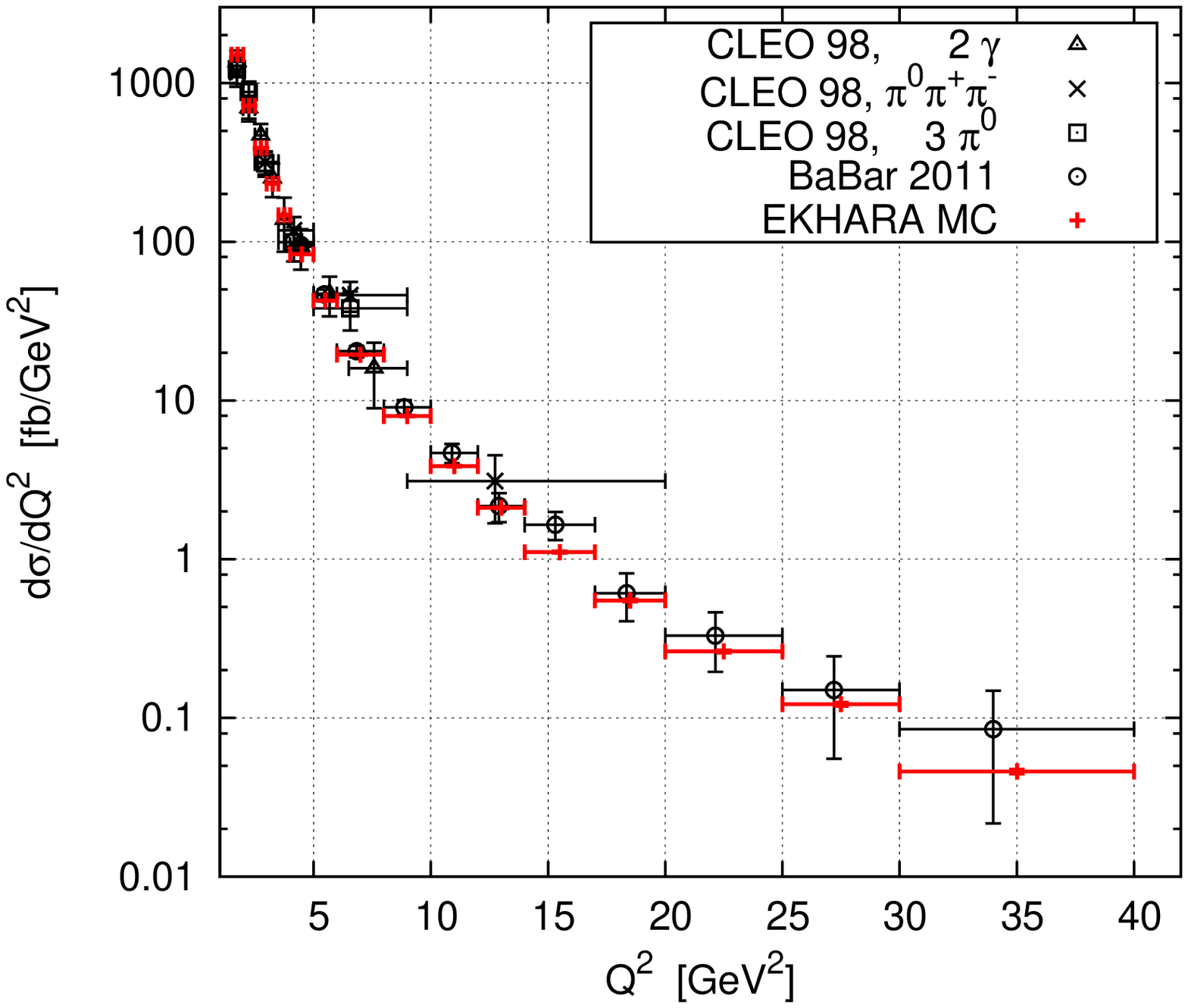}
     }
\resizebox{0.48\textwidth}{!}{%
     \includegraphics{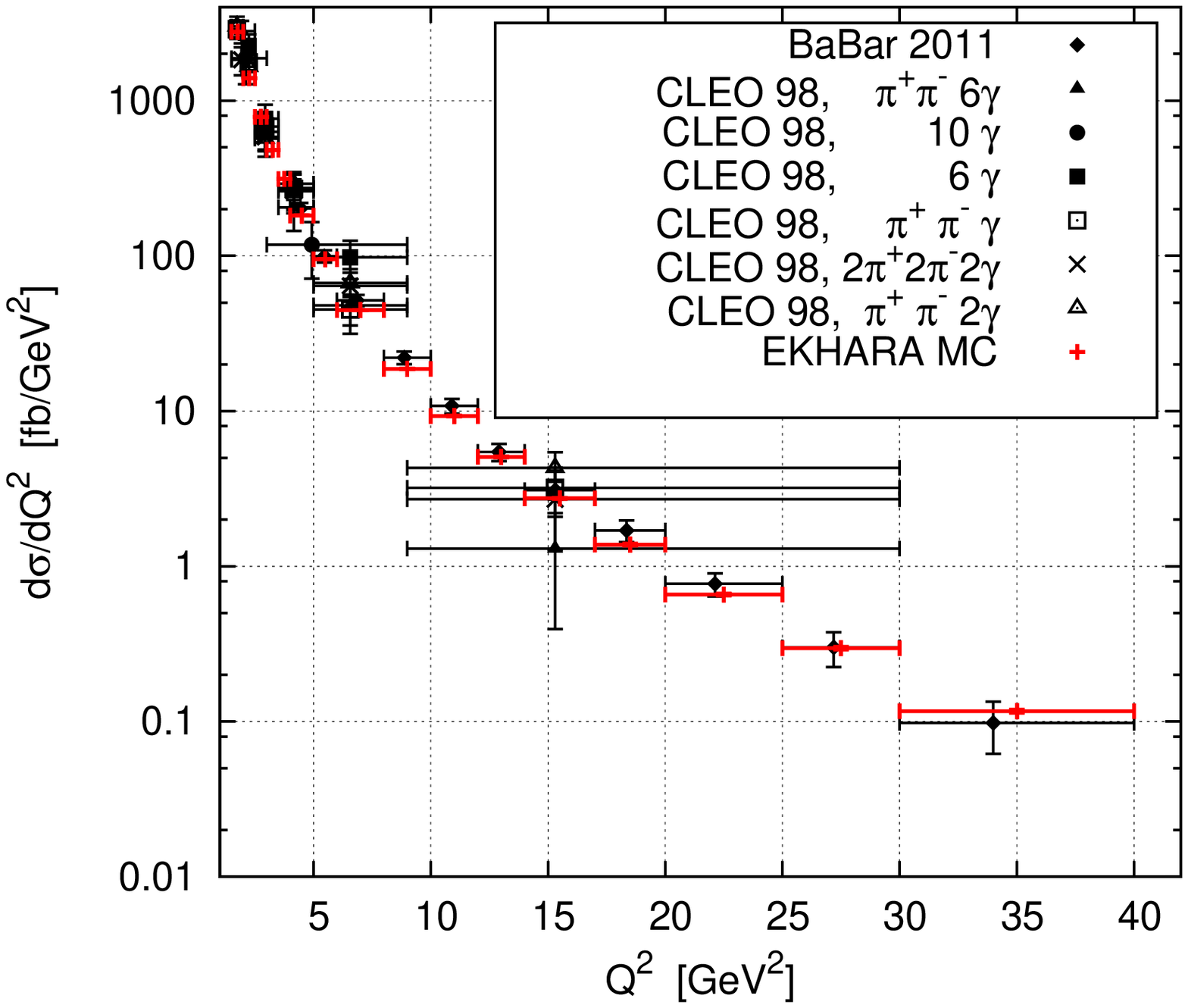}
     }
\end{center}
\caption{The cross section ${d\sigma}/{d Q^2}$
for the process $e^+e^- \to e^+e^-\eta$ (left) and 
$e^+e^- \to e^+e^-\eta^\prime$ (right)
 compared to CLEO~\cite{Gronberg:1997fj} and BaBar~\cite{:2011hk} data.}
 \label{fig:cs-eta}
\end{figure*}

We simulate the cross sections ${d\sigma}/{d Q^2}$
for the process $e^+e^- \to e^+e^-\mathcal{P}$
and compare it with existing ``single-tag'' data from the 
  CELLO~\cite{Behrend:1990sr}, CLEO~\cite{Gronberg:1997fj} 
  and BaBar~\cite{Aubert:2009mc,:2011hk} experiments.
In a single-tag experiment,
the ``tagged'' lepton fixes the value of $Q^2 = -t_1$
and the 4-momentum squared of the ``untagged'' lepton $t_2 = -q_2^2$ 
is kinematically restricted nearby zero.
For example, in the BaBar experiment, the actual thresholds 
for $q_2^2$ are $0.18$~GeV$^2$ for pions~\cite{Aubert:2009mc}
and $0.38$~GeV$^2$ for $\eta$ and $\eta^\prime$~\cite{:2011hk}
due to the imposed event selection.
The experimental ${d\sigma}/{d Q^2}$ is given within these cuts,
and, therefore, the simulated ${d\sigma}/{d Q^2}$ is computed within
the similar event selection.
As expected, a good agreement between the generator predictions
and the data is observed, see Figs.~\ref{fig:cs-pi0},~\ref{fig:cs-eta}.
%


%
%

An important note here is in order.
The values of the ${d\sigma}/{d Q^2}$ are the primary results of the experiment.
The form factor $F_{\mathcal{P}}(Q^2,0)$ is calculated then on the basis of 
the measured ${d\sigma}/{d Q^2}$ and the simulation.
It is known that the model dependence in the simulation leads to the uncertainty 
in the form factor, which is ``measured'' in this way.
In the BaBar analyses, the corresponding uncertainties in cross section
are estimated to be at the level of $3.5~\%$ for pions~\cite{Aubert:2009mc}
and $4.6~\%$ for $\eta$ and $\eta^\prime$~\cite{:2011hk}
(based on the simulation with the $q_2^2$--dependent
and $q_2^2$--independent form factors).
As stressed in~\cite{Aubert:2009mc}, this uncertainty 
is very sensitive to the actual $q_2^2$ cut.

\begin{figure}
\begin{center}
\resizebox{0.42\textwidth}{!}{%
     \includegraphics[angle=270]{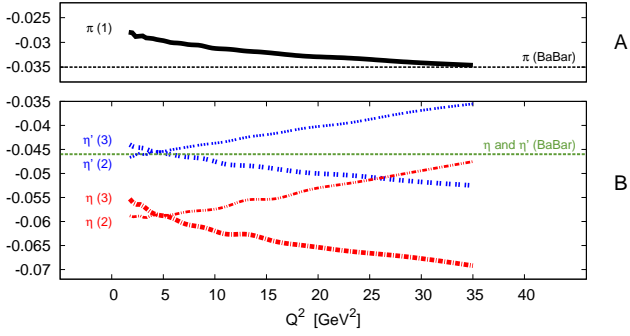}
     }
\end{center}
\caption{The relative difference of the cross sections
$({d\sigma[full]} - {d\sigma[approx]})/{d\sigma[full]}$
for the process $e^+e^- \to e^+e^-\pi^0$ (A) and $e^+e^- \to e^+e^-\eta^{(\prime)}$ (B).
The approximate simulation ($d\sigma[approx]$)
ignores the form factor dependence on the momentum transfer ($q_2^2$) 
to the untagged lepton.
The full simulation ($d\sigma[full]$) accounts for the virtuality of both photons 
in the form factor.
The following cuts are used in the simulation: \\
(1) $|q_2^2| < 0.18$~GeV$^2$~\cite{Aubert:2009mc}; \\
(2) $|\cos \theta^\ast_{ e \mathcal{P} }|>0.99$ and $|q_2^2| < 0.6$~GeV$^2$~\cite{:2011hk}; \\
(3) $|q_2^2| < 0.38$~GeV$^2$~\cite{:2011hk}. \\
The lines denoted by~(BaBar) show estimates for the relative difference,
as given in the BaBar papers:
for $\pi^0$ --- in~\cite{Aubert:2009mc}, 
for $\eta$ and $\eta^\prime$ --- in~\cite{:2011hk}.
}
 \label{fig:cs-relative}
\end{figure}

Recently, the effects of the $q_2^2$ cut were emphasized on 
the level of the form factor considerations~\cite{Lichard:2010ap}.
The Monte Carlo generator in hand allows us to perform a more
conclusive study, namely to investigate the magnitude of the
cross section uncertainties, discussed above.
Similarly to the method used in the BaBar~\cite{Aubert:2009mc,:2011hk} analyses we
perform two simulations: the first (${d\sigma[full]}/{dQ^2}$) 
with the exact form factor
$F_{\gamma^\ast\gamma^\ast \mathcal{P}}(t_1, t_2)$
and the second (${d\sigma[approx]}/{dQ^2}$)
with the approximated form factor 
$F_{\gamma^\ast\gamma^\ast \mathcal{P}}(t_1, t_2)
\approx F_{\gamma^\ast\gamma \mathcal{P}}(t_1, 0)$,
i.e., neglecting the momentum transfer to the untagged lepton in the form factor.
The relative difference of the corresponding cross sections 
is then plotted in Fig.~\ref{fig:cs-relative}.
Our estimations for the uncertainty
are in a rough agreement with that of BaBar~\cite{Aubert:2009mc,:2011hk}.
However, in contrast to the estimate of BaBar,
a dependence of this uncertainty on $Q^2$ is observed
in our simulation.
If this effect is not accounted for in the data it might
result in inducing a fake $Q^2$ dependence of the form factor.

In order to investigate the impact of the event selection 
on the error estimate, we perform the simulation for $\eta$ and $\eta^\prime$ mesons
with the direct cut on the second (untagged) invariant ($q_2^2$) 
and separately with the cut on the angle between the initial 
and final untagged lepton ($|\cos\theta^\ast_{ e \mathcal{P} }| > 0.99$),
which effectively induces the cut on $q_2^2$~\cite{Aubert:2009mc,:2011hk}.
From Fig.~\ref{fig:cs-relative} we see that the error estimate 
and its $Q^2$ dependence is very sensitive to the event selection.
%



\section{The limits of the form factors}
\label{sec:limits}
\subsection{The high-$Q^2$ limit of the form factors}
\label{sec:limit:high}

The issue of the asymptotic behavior of the form factors
usually deserves an attention.
In our approach, the  high-$Q^2$ limits (${t\rightarrow -\infty}$) are the following.
In case of the one octet ansatz, we obtain
from Eqs.~(\ref{eq:ffpi_3octet}),~(\ref{eq:ffeta_3octet}) and~(\ref{eq:ffetapr_3octet}):
\begin{widetext}
\bgea
    F_{\gamma^\ast \gamma \pi^0}(t,0) &=& \frac{1}{8 \pi^2 f_\pi}\frac{1}{t} (M_\rho^2 + M_\omega^2)
+ \mathcal{O} \Bigl(\frac{1}{t^2}  \Bigr)
 \label{eq:ff_limit:pi_1}   
 \ ,
\enea
\bgea
    F_{\gamma^\ast \gamma^\ast \pi^0}(t,t) &=& \frac{1}{4 \pi^2 f_\pi}\frac{1}{t^2} M_\rho^2 M_\omega^2
 + \mathcal{O} \Bigl(\frac{1}{t^3} \Bigr)   
 \label{eq:ff_limit:pi_2}   
 \ ,
\enea
\bgea
    &&F_{\gamma^\ast \gamma \eta}(t,0) = \frac{1}{8 \pi^2 f_\pi} \frac{1}{t}
    \bigl(3 C_q M_\rho^2 + \frac{1}{3} C_q M_\omega^2 -\frac{2\sqrt{2}}{3} C_s M_\phi^2 \bigr) 
+ \mathcal{O} \Bigl(\frac{1}{t^2} \Bigr)
 \label{eq:ff_limit:eta_1}   
 \ ,
\enea
\bgea
    &&F_{\gamma^\ast \gamma^\ast \eta}(t,t) = \frac{1}{8 \pi^2 f_\pi}\frac{1}{t^2}
    \bigl(3 C_q M_\rho^4 + \frac{1}{3} C_q M_\omega^4 -\frac{2\sqrt{2}}{3} C_s M_\phi^4 \bigr) 
+ \mathcal{O} \Bigl(\frac{1}{t^3} \Bigr)
 \label{eq:ff_limit:eta_2}   
 \ .
\enea
In case of the two octets ansatz, we obtain
\bgea
    F_{\gamma^\ast \gamma \pi^0}(t,0) &=& \frac{4\sqrt{2}}{3 f_\pi}\frac{1}{t} 
    \left[h_{V_1}f_{V_1} (M_\rho^2 + M_\omega^2)\right.
        + \left.h_{V_2}f_{V_2} (M_{\rho^\prime}^2 + M_{\omega^\prime}^2)\right]
+ \mathcal{O} \Bigl(\frac{1}{t^2} \Bigr)
 \label{eq:ff_limit:pi_3}   
 \ ,
\enea
\bgea
    F_{\gamma^\ast \gamma^\ast \pi^0}(t,t) &=& \frac{8\sqrt{2}}{3 f_\pi}\frac{1}{t^2} 
    \left[h_{V_1}f_{V_1} M_\rho^2 M_\omega^2\right.
        + \left.h_{V_2}f_{V_2} M_{\rho^\prime}^2 M_{\omega^\prime}^2\right]
+ \mathcal{O} \Bigl(\frac{1}{t^3} \Bigr)
 \ ,
 \label{eq:ff_limit:pi_4}   
\enea
\bgea
\!\!\!\!\!\!\!\!\!\!\!
    &&\!F_{\gamma^\ast \gamma \eta}(t,0) = \frac{4\sqrt{2}}{3 f_\pi} \frac{1}{t}
    \!\!\!\left[\!h_{V_1}f_{V_1} \bigl(3 C_q M_\rho^2 + \frac{1}{3} C_q M_\omega^2 -\frac{2\sqrt{2}}{3} C_s M_\phi^2 \bigr)\right. 
       \! + \! \left.h_{V_2}f_{V_2} \bigl(3 C_q M_{\rho^\prime}^2 + \frac{1}{3} C_q M_{\omega^\prime}^2 -\frac{2\sqrt{2}}{3} C_s M_{\phi^\prime}^2 \bigr)\!\right]
\!\!\!+ \!\mathcal{O} \Bigl(\frac{1}{t^2} \Bigr)
 \label{eq:ff_limit:eta_3}   
   \;,\ \  \  \  \  \  \  \ 
\enea
\bgea
\!\!\!\!\!\!\!\!\!\!\!
    &&\!F_{\gamma^\ast \gamma^\ast \eta}(t,t) = \frac{8\sqrt{2}}{3 f_\pi}\frac{1}{t^2}
    \!\!\!\left[\!h_{V_1}f_{V_1} \bigl(3 C_q M_\rho^4 + \frac{1}{3} C_q M_\omega^4 -\frac{2\sqrt{2}}{3} C_s M_\phi^4 \bigr) \right.
       \! + \! \left.h_{V_2}f_{V_2} \bigl(3 C_q M_{\rho^\prime}^4 + \frac{1}{3} C_q M_{\omega^\prime}^4 -\frac{2\sqrt{2}}{3} C_s M_{\phi^\prime}^4 \bigr)\!\right]     
\!\!\!+ \!\mathcal{O} \Bigl(\frac{1}{t^3} \Bigr)
 \label{eq:ff_limit:eta_4}   
\;. \  \  \  \  \  \  \  \ 
\enea
\end{widetext}
The limits for the $\eta^\prime$ form factor 
can be obtained from the above formulae 
according to~(\ref{eq:ffetapr_3octet}).

The expressions~(\ref{eq:ff_limit:pi_3}) and~(\ref{eq:ff_limit:eta_3})
guide the high-$Q^2$ behavior of the form factors
measured in single-tag experiments
(shown in Figs.~\ref{fig:ff-pi0},~\ref{fig:ff-eta},~\ref{fig:ff-etaP}).
We see that in our approach the asymptotic 
value of $Q^2|F_{\mathcal{P}}(Q^2,0)|$
depends not only on the mixing parameters and decay constants
but also on the masses of the vector resonances.
Numerically, for $\pi^0$ transition form factor,
the value of $Q^2|F_{\pi^0}(Q^2,0)|$ in our approach
with two octets is very close to that 
of the Brodsky-Lepage~\cite{Lepage:1980fj} high-$Q^2$ limit 
$Q^2|F_{\pi^0}(Q^2,0)| \to 2 f_\pi$
shown as a bold solid line (BL) in Fig.~\ref{fig:ff-pi0}.

The perturbative QCD prediction for the asymptotic 
of the $\eta$ and $\eta^\prime$ form factors
is often given in a simple approach in terms of the
parameters $f_\eta=0.0975$~GeV and
$f_{\eta^\prime}=0.0744$~GeV~\cite{Gronberg:1997fj,Dorokhov:2009jd}:
$Q^2|F_{\eta^{(\prime)}}(Q^2,0)| \to 2 f_{\eta^{(\prime)}}$.
These values are shown as bold solid line (BL)
in Figs.~\ref{fig:ff-eta},~\ref{fig:ff-etaP}
and in this case one can notice no coincidence 
with the values given by~(\ref{eq:ff_limit:eta_3}).
Sometimes it is also called the Brodsky-Lepage limit,
with a reference to~\cite{Brodsky:1981rp}.
However, we would like to remark that 
in~\cite{Brodsky:1981rp} the $SU(3)$ flavor
breaking effects are not considered and the assumed $\eta-\eta^\prime$
mixing may be not consistent with modern data.
An attempt to interpret the results of~\cite{Brodsky:1981rp}
by means of 
$f_{\eta^\prime}=0.0744$~GeV
is in tension with the data for $\eta^\prime$ form factor, as
one can see in Fig.~\ref{fig:ff-etaP} 
and as noticed in~\cite{Dorokhov:2009jd}.

In principle, there are other ways to apply the formulae
of~\cite{Brodsky:1981rp} to the form factors of physical states
$\eta$ and $\eta^\prime$~\cite{Ametller:1991jv}.
For example, the limit for the $\eta$ and $\eta^\prime$ 
transition form factors can be calculated 
according to the two-angle $\eta-\eta^\prime$ 
mixing scheme~\cite{Feldmann:1999uf,Kroll:2010bf}.
The latter values are shown as a shaded box (FK)
in Figs.~\ref{fig:ff-eta},~\ref{fig:ff-etaP} (green online).
Notice, that numerically this limit for $\eta$  meson 
is very close to that of BL approach and 
also to the value given by our model,
however for $\eta^\prime$ all the three values are different.

\subsection{The slope of the form factor at the origin}
\label{sec:slope}

Sometimes it is convenient to define the so-called 
slope of the transition form factor at the origin
$a_{\mathcal{P}}$ (``slope parameter''):
\bgea
a_{\mathcal{P}} &\equiv& \left.\frac{1}{F_{\gamma^\ast\gamma^\ast\mathcal{P}}(0,0)}
\frac{d\ F_{\gamma^\ast\gamma^\ast\mathcal{P}}(t,0)}{d\ x}\right|_{t=0},
\enea
where $x\equiv t/m^2_{\mathcal{P}}$.
Notice that being defined this way, $a_{\mathcal{P}}$
is a dimensionless quantity,
which is related to the effective region
of the $\gamma\gamma^\ast\mathcal{P}$ interaction
$\langle{r^2_{\mathcal{P}}}\rangle = 6 a_{\mathcal{P}}/m^2_{\mathcal{P}}$. 
The average experimental value for $a_{\pi}$ 
listed in PDG~\cite{Nakamura:2010zzi}
(linear coefficients of the $\pi^0$ electromagnetic 
form factor) 
is mainly driven not by a direct measurement,
but by an extrapolation done in Ref.~\cite{Behrend:1990sr}.
The direct measurements of $a_{\pi}$
are less precise~\cite{Farzanpay:1992pz,MeijerDrees:1992qb}.
The experimental knowledge of $a_{\eta}$ is much better
and recently the new experiments contributed: 
MAMI-C~\cite{Berghauser:2011zz} and NA60~\cite{:2009wb,Usai:2011zz}. 
For $a_{\eta^\prime}$ we were not able to find a result of
a direct measurement.

From Eqs.~(\ref{eq:ffpi_3octet}),~(\ref{eq:ffeta_3octet}) and~(\ref{eq:ffetapr_3octet})
one obtains the following model prediction for the slope parameters:
\bgea
\label{eq:slope:pi}
a_{\pi} &=&
\frac{16\sqrt{2}\pi^2m^2_{\pi}}{N_C}
\sum_{i=1}^{n}{
     h_{V_i} f_{V_i} \left( \frac{1}{M^2_{\rho_i}} + \frac{1}{M^2_{\omega_i}}\right)
    }
    \;,
\enea
\bgea
a_{\eta} &=&
\frac{16\sqrt{2}\pi^2 m^2_{\eta}}{N_C}
\left( \frac{5}{3}C_q - \frac{\sqrt{2}}{3}C_s \right)^{-1}
\nn\\
&&\!\!\!\!\!\!\times
 \sum_{i=1}^{n}{
     h_{V_i} f_{V_i} \left( \frac{3C_q}{M^2_{\rho_i}} + \frac{C_q}{3M^2_{\omega_i}} - \frac{2\sqrt{2}C_s}{3 M^2_{\phi_i}}\right)
    }
    \;,
\label{eq:slope:eta}
\enea
\bgea
a_{\eta^\prime} &=&
\frac{16\sqrt{2}\pi^2 m^2_{\eta^\prime}}{N_C}
\left( \frac{5}{3}C_q^\prime + \frac{\sqrt{2}}{3}C_s^\prime \right)^{-1}
\nn\\
&&\!\!\!\!\!\!\times
 \sum_{i=1}^{n}{
     h_{V_i} f_{V_i} \left( \frac{3C_q^\prime}{M^2_{\rho_i}} + \frac{C_q^\prime}{3M^2_{\omega_i}} + \frac{2\sqrt{2}C_s^\prime}{3 M^2_{\phi_i}}\right)
    }
    \;.
\label{eq:slope:etaP}
\enea

The numerical values for $a_{\mathcal{P}}$ 
are listed in Table~\ref{table:slope}.
On its basis we conclude that there is a reasonable agreement 
between model predictions and experiments.

We would like to remark that in the limit of the 
equal masses for vector resonances within the octet, 
Eqs.~(\ref{eq:slope:pi}),~(\ref{eq:slope:eta}),~(\ref{eq:slope:etaP})
lead to the following relation between the slope parameters:
$a_{\pi}/m^2_{\pi} = a_{\eta}/m^2_{\eta} = a_{\eta^\prime}/m^2_{\eta^\prime}$.

\begin{table*}
\caption{
Model prediction for the slope parameters
$a_{\mathcal{P}}$ and two most recent experimental
values.
The ``2 octets'' column is calculated with 
the parameter values given by our global fit. 
The first error in experimental value is due to
statistics and the second one is systematics.
}
\label{table:slope}
\begin{ruledtabular}
\begin{tabular}{lllll}
& \multicolumn{1}{c}{1 octet} & \multicolumn{1}{c}{2 octets} & \multicolumn{2}{c}{experiments}\\
\hline\noalign{\smallskip}
$a_{\pi}$         & 0.03003(1) & 0.02870(9) & 0.026(24)(48)~\text{\cite{Farzanpay:1992pz}}& 0.025(14)(26)~\text{\cite{MeijerDrees:1992qb}}\\
$a_{\eta}$        & 0.546(9) & 0.521(2) & 0.576(105)(39)~\text{\cite{Berghauser:2011zz}}& 0.585(18)(13)~\text{\cite{Usai:2011zz}}\\
$a_{\eta^\prime}$ & 1.384(3) & 1.323(4) & \text{---}& \text{---}\\
\end{tabular}
\end{ruledtabular}
\end{table*}

\section{Summary}
\label{sec:summary}

Using the scheme of the $\eta$--$\eta^\prime$ mixing
with two decay parameters ($f_0$ and $f_8$) 
and two mixing angles ($\theta_0$, $\theta_8$)~\cite{Feldmann:1998vh,Feldmann:1999uf}
and following the approach of chiral effective theory with resonances~\cite{Ecker:1989yg,Ecker:1988te,Prades:1993ys},
we derive the expressions for the two-photon transition form 
factors of the $\mathcal{P}=\pi^0$, $\eta$, $\eta^\prime$ mesons.
The tree-level contributions within this effective field theory
approach are considered.
  For the case of the one octet ansatz there are no free parameters
  and we obtain the model prediction for the form factors.
  For the case of the two octet ansatz the model parameter
  is fitted to the data.
We find that the two octet calculation is consistent with the bulk 
 of available data.
The high-$Q^2$ limits of the form factors
in our approach are compared to those of
the Brodsky and Lepage~\cite{Lepage:1980fj,Gronberg:1997fj,Dorokhov:2009jd}
and to those of Feldmann and Kroll~\cite{Feldmann:1999uf,Kroll:2010bf}.
The slope of the transition form factor at the origin,
$a_{\mathcal{P}}$, is calculated and compared to available data.
A reasonable agreement between model predictions and experiments is found.

The obtained form factors are implemented in the EKHARA Monte Carlo
generator.
As a test of the generator, the cross-section ${d\sigma}/{d Q^2}$ is simulated 
for the process $e^+e^- \to e^+e^-\mathcal{P}$ and compared to data
using the event selections,
which mimic the ``single-tag'' experimental conditions. 
%
%

Using the Monte Carlo simulation we investigate the
impact of neglecting the momentum transfer to the untagged lepton 
($t_2$) on the cross section and form factor measurements.
The uncertainty in the visible cross section due to the
simplification of the form factor 
$F_{\gamma^\ast\gamma^\ast\mathcal{P}}(t_1,t_2) \approx F_{\gamma^\ast\gamma^\ast\mathcal{P}}(t_1,0)$
is estimated for the phase space cuts 
similar to the experimental ones.

Due to very small number of free parameters 
and good agreement with data, the approach presented
in this work is a good starting point
for further model adjustments, e.g., for including the
$SU(3)$ flavor symmetry breaking in the couplings.
Using the developed generator one will be able to study, 
e.g., a possible manifestation of such effects 
in the cross section ${d\sigma}/{d Q^2}$ within 
a realistic phase space cuts.

\section*{Acknowledgments}
We would like to kindly acknowledge our colleagues from
experimental collaborations, whose feedback about the EKHARA generator
and the physics case of $e^+e^- \to e^+e^-\mathcal{P}$ gave a motivation
of the current research. 
We greatly profited from discussions
with Danilo Babusci, Achim Denig, Dario Moricciani, Matteo Mascolo, 
Elisabeta Prencipe and Graziano Venanzoni.
We also thank Fred Jegerlehner and Andreas Nyffeler for discussions.


This research was partly supported by Marie Curie Intra
European Fellowship within the 7th European Community Framework Programme
(FP7-PEOPLE-2009-IEF),
by Polish Ministry of Science and High Education
from budget for science for years 2010-2013: grant number N~N202~102638,
by National Academy of Science
of Ukraine under contract $50/53$---$2011$,
%
by Sonderforschungsbereich SFB1044 of the Deutsche Forschungsgemeinschaft,
by Spanish Consolider Ingenio 2010 Programme
CPAN (CSD2007-00042) as well by MEC (Spain) under Grants {FPA2007-60323},
{FPA2011-23778}.
This work is a part of the activity of the ``Working Group on Radiative
Corrections and Monte Carlo Generators for Low 
Energies''(\href{http://www.lnf.infn.it/wg/sighad/}{{\tt http://www.lnf.infn.it/wg/sighad/}})~\cite{Actis:2010gg}.

\appendix

\section{Formalism}
\label{sec:formalism}

The lightest pseudoscalar mesons 
are supposed to play a role of the 
(pseudo-)Nambu-Goldstone boson fields of spontaneously
$G=SU(3)_L\times SU(3)_R$ to $H=SU(3)_V$ broken symmetry. 
To introduce the physical states $\eta$ and $\eta^\prime$ 
we choose the scheme with two mixing angles ($\theta_0$, $\theta_8$), see~\cite{Feldmann:1998vh,Feldmann:1999uf}. 
The nonet of the pseudoscalar mesons reads 
\bgea
\label{eq:ps-nonet}
u &=& \\
 && \!\!\!\!\! exp\!\left\{\!\frac{i}{\sqrt{2}f_\pi}
\!\left(
\!\begin{array}{ccc}
 \frac{\pi^0 + C_{q}\eta + C_{q}^\prime \eta^\prime}{\sqrt{2}} & \pi^+  & \frac{f_\pi}{f_K} {K^+ } \\
 \pi^- & \hspace{-0.18in} \frac{-\pi^0+ C_{q}\eta + C_{q}^\prime \eta^\prime}{\sqrt{2}} & \frac{f_\pi}{f_K} {K^0} \\
 \frac{f_\pi}{f_K} {K^-} & \frac{f_\pi}{f_K} {\bar{K}^0} & \hspace{-0.06in} - C_{s}\eta + C_{s}^\prime \eta^\prime
\end{array}
\!\right)\!\right\}
\nn
\enea
where $f_{\pi}$ and $f_{K}$ are the pion and kaon decay constants
and the following notation is used
\bgea
\label{eq:eta-coefficients}
C_q &\equiv& \frac{f_\pi}{\sqrt{3} \cos(\theta_8-\theta_0)}
\left( \frac{1}{f_8}\cos\theta_0 - \frac{1}{f_0}\sqrt{2} \sin\theta_8 \right),
\nn
\\
C^\prime_q &\equiv& \frac{f_\pi}{\sqrt{3} \cos(\theta_8-\theta_0)}
\left(\frac{1}{f_0} \sqrt{2} \cos\theta_8 + \frac{1}{f_8}\sin\theta_0 \right),
\nn
\\
C_s &\equiv& \frac{f_\pi}{\sqrt{3} \cos(\theta_8-\theta_0)}
\left( \frac{1}{f_8}\sqrt{2}\cos\theta_0 +  \frac{1}{f_0}\sin\theta_8 \right),
\nn
\\
C^\prime_s &\equiv& \frac{f_\pi}{\sqrt{3} \cos(\theta_8-\theta_0)}
\left(\frac{1}{f_0}\cos\theta_8 -  \frac{1}{f_8}\sqrt{2}  \sin\theta_0 \right)
.
\enea
Fixing the angles $\theta_0$,$\theta_8$ and constants $f_0$,$f_8$~\cite{Feldmann:1998vh,Feldmann:1999uf}
\bgea
\label{eq:mix_angles}
\theta_8 &=& -21.2^\degree \pm 1.6^\circ, \quad  \theta_0 =  -9.2^\degree \pm 1.7^\circ ,
\nn \\
f_8 &=& (1.26 \pm 0.04) f_\pi, \quad  f_0 = (1.17 \pm 0.03) f_\pi,
\enea
and taking $f_{\pi} = 92.4$~MeV, one obtains $C_q \approx 0.720$,
$C_s \approx 0.471$, $C_q^\prime \approx 0.590$ and
$C_s^\prime \approx 0.576$.
Notice that accordingly to notation~(\ref{eq:ps-nonet}) 
the couplings of the $\eta^\prime$ are easily related to
those of the $\eta$ meson by means of substitution
\bgea
 C_q & \to &  C_q^\prime, \nn \\
 C_s & \to & -C_s^\prime.
\enea
Obviously, this pattern also holds in the expressions 
for the form factors in our approach.

At the lowest order the Wess-Zumino-Witten Lagrangian~\cite{Wess:1971yu,Witten:1983tw}, 
that describes the interaction of pseudoscalar mesons with two photons, 
can be written down in the terms of the  physical fields as
\bgea
\label{eq:WZW-lagr}
\mathcal{L}_{\gamma\gamma P} 
=
 &-& \frac{e^2 N_c}{24 \pi^2 f_\pi}
\epsilon^{\mu\nu\alpha\beta} \partial_\mu B_\nu
\partial_\alpha B_\beta   
\nn\\
&\times& \left[\pi^0 + \eta \left( \frac{5}{3} C_q - \frac{\sqrt{2}}{3} C_s\right)\right.
\nn\\
&&\left.+ \eta^\prime \left( \frac{5}{3} C_q^\prime + \frac{\sqrt{2}}{3} C_s^\prime\right)
  \right] 
,
\enea
where $N_c =3$ is the number of quark colors
and the electromagnetic field is denoted by $B_\nu$.

Assuming the $SU(3)$ symmetry for the 
coupling constants of the vector mesons, 
the $\gamma V$ interaction is written as
\bgea
\mathcal{L}_{\gamma V} &=&  - e \sum_{i=1}^{n} f_{V_i}  \partial_\mu B_\nu \bigl(
\tilde{\rho}_i^{\mu\nu} + \frac{1}{3}\tilde{\omega}_i^{\mu\nu} -
\frac{\sqrt{2}}{3}\tilde{\phi}_i^{\mu\nu} \bigr)
 \label{eq:vector_gamma_V}
\enea
where we have summed over octets of the vector mesons, $\tilde{V}_{\mu \nu} \equiv \partial_\mu V_\nu -
\partial_\nu V_\mu$, $f_{V_i}$ is the (dimensionless) coupling for
the vector representation of the spin-1 fields for a fixed octet.

The Lagrangians that describes vector-photon-pseudoscalar and two vector mesons interactions with pseudoscalar~\cite{Prades:1993ys} in the terms of the physical fields read 
\bgea
\mathcal{L}_{V\gamma \mathcal{P}}&=& -\sum_{i=1}^{n}\frac{4\sqrt{2} e h_{V_i}}{3
f_\pi}\epsilon_{\mu\nu\alpha\beta} \partial^\alpha B^\beta \biggl[ (
\rho_i^\mu  +3\omega_i^{\mu} )
\partial^\nu \pi^0\nn 
\\ 
\label{lagr_vgp}
&&
 + \bigl[ (3 \rho_i^{\mu} + \omega_i^{\mu})C_q + 2 \phi_i^{\mu} C_s \bigr] \partial^\nu \eta
\nn\\ 
&&
 + \bigl[ (3 \rho_i^{\mu} + \omega_i^{\mu})C_q^\prime - 2 \phi_i^{\mu} C_s^\prime \bigr] \partial^\nu \eta^\prime
 \biggr],
\enea
\bgea
\nn
\mathcal{L}_{V V \mathcal{P}}&=& - \sum_{i=1}^{n} \frac{4\sigma_{V_i}}{f_\pi}\epsilon_{\mu\nu\alpha\beta}
\biggl[
\pi^0 \partial^\mu \omega_i^\nu \partial^\alpha \rho_i^{\beta}
\\
&&
 +  \eta \bigl[  (\partial^\mu\rho_i^{\nu} \partial^\alpha
\rho_i^{\beta}+
\partial^\mu \omega_i^{\nu} \partial^\alpha \omega_i^{\beta} )
  \frac{1}{2}\,C_q
\nn\\
&&
 - \partial^\mu \phi_i^{\nu}\partial^\alpha \phi_i^{\beta}
\frac{1}{\sqrt{2}} \, C_s \bigr] 
\nn\\
&&
 +  \eta^\prime \bigl[  (\partial^\mu\rho_i^{\nu} \partial^\alpha
\rho_i^{\beta}+
\partial^\mu \omega_i^{\nu} \partial^\alpha \omega_i^{\beta} )
  \frac{1}{2}\,C_q^\prime
\nn\\
&&
 + \partial^\mu \phi_i^{\nu}\partial^\alpha \phi_i^{\beta}
\frac{1}{\sqrt{2}} \, C_s^\prime \bigr] 
\biggr]
,
\label{lagr_vvp}
\enea
where $h_{V_i}$ and $\sigma_{V_i}$ are the corresponding (dimensionless) 
coupling constants for a given $i$-th octet. 
For simplicity we neglect any mixing between the octets.

%
%



\end{document}